\documentstyle[12pt,aps,epsf]{revtex}

\begin{document}




\title{
\hfill \parbox[t]{2 in} {\rm \small
ORNL-CTP-95-02\\RAL-94-106\\hep-ph/9501405}
\vskip 1.3 cm
Hybrid and Conventional Mesons in the Flux Tube Model:
Numerical Studies and their Phenomenological Implications
}

\vskip 1cm
\author{T. Barnes}
\address{
Computational and Theoretical Physics Group,
Oak Ridge National Laboratory\\
Oak Ridge, TN 37831-6373, USA\\
and\\
Department of Physics and Astronomy,
University of Tennessee\\
Knoxville, TN 37996-1200, USA
}

\vskip 1cm
\author{F.E. Close}
\address{
Daresbury Rutherford Appleton Laboratory\\
Chilton, Didcot, OXON OX11 0QX, England
}

\vskip 1cm
\author{E.S. Swanson}
\address{
Department of Physics,
North Carolina State University\\
Raleigh, NC 27695-8202, USA\\
}
\date{January 1995}
\maketitle

\vspace{0.5cm}
\begin{center}
{\bf Abstract}
\end{center}
\vspace {.2cm}

\begin{abstract}
We present results from analytical and numerical studies of
a flux tube model of hybrid mesons. Our numerical results use a Hamiltonian
Monte Carlo algorithm and so improve on previous analytical treatments,
which assumed small flux tube oscillations
and an adiabatic separation of quark and
flux tube motion. We find that the small oscillation approximation is
inappropriate for typical hadrons and that the
hybrid mass is
underestimated by the adiabatic approximation.
For physical parameters
in the ``one-bead" flux tube model we estimate the lightest hybrid masses
(${}_\Lambda L  = {}_1 P$ states) to
be
1.8-1.9~GeV for $u\bar u$ hybrids,
2.1-2.2~GeV for $s\bar s$ and
4.1-4.2~GeV for $c\bar c$.
We also determine masses of conventional $q\bar q$ mesons
with $L=0$ to $L=3$ in this model, and confirm good agreement
with experimental $J$-averaged multiplet masses.
Mass estimates are also given for hybrids with higher
orbital and flux-tube excitations.
The gap from the lightest hybrid level
(${}_1P$)
to the first hybrid orbital excitation
(${}_1D$)
is predicted to be
$\approx 0.4$~GeV for light quarks $(q=u,d)$ and
$\approx 0.3$~GeV for $q=c$.
Both ${}_1P$ and ${}_1D$
hybrid multiplets contain the exotics $1^{-+}$ and $2^{+-}$; in addition
the ${}_1P$ has a $0^{+-}$ and the ${}_1D$ contains a $3^{-+}$.
Hybrid mesons with doubly-excited flux tubes
are also considered.
The implications of our results for spectroscopy are discussed,
with
emphasis on charmonium hybrids, which may be accessible at
facilities such as BEPC, KEK, a Tau-Charm Factory, and in
$\psi$ production at hadron colliders.

\end{abstract}

\newpage

\section{Introduction}

The QCD Lagrangian contains quarks and gluons and the successes of
perturbative QCD confirm their existence as dynamical degrees of freedom.
The behavior of QCD in the strongly
interacting low-energy regime, ``nonperturbative QCD", is
less well understood.
Studies using lattice gauge theory have confirmed the
presence of confinement and give spectra for conventional mesons and
baryons that are in reasonable agreement with experiment
\cite{LGTrev}, but the status of
gluonic hadrons in the spectrum has remained obscure.

It is possible that this is now about to change. Candidates for gluonic
hadrons have recently been reported which have much in common
with theoretical expectations.
There are various lattice
predictions for the masses of glueballs; the most reliable is presumably
for the glueball ground state, which is expected to
be a scalar with a mass near $1.5$-$1.7$~GeV~\cite{LGTrev}.
A candidate for the scalar glueball has been reported at 1520 MeV
 by the Crystal Barrel collaboration at LEAR \cite{CBar} and may also be
evident
in central production by NA12/2\cite{NA12} at CERN.
Possible evidence for a
$1^{-+}$ light exotic hybrid candidate has been reported in
$\rho\pi$ and $f_2\pi$ at about 1775 MeV \cite{Condo} in $\eta\pi$ and
especially $\eta'\pi$ at $\sim 1.6$ GeV by VES \cite{VES}, and in
$f_1\pi$ \cite{f1pi} with a resonant phase
in the region 1.6-2.2 GeV,
with production and decay characteristic similar to theoretical expectations
for
``hybrid" states.
A light $1^{-+}$ signal in $\eta\pi$
reported by GAMS near 1.4 GeV \cite{GAMS} has been withdrawn, although KEK
\cite{KEK} reports
a resonant $1^{-+}$ amplitude with a mass and width similar to the $a_2(1320)$.
Another possibility is that
the surprisingly large $\psi '$ production at the
Tevatron \cite{Tev} may be due to
the formation and decay of metastable hybrid charmonium\cite{C94}.

In view of the discovery of these candidates
for gluonic hadrons it is appropriate
to investigate the theoretical models for
these states more carefully, to see if the
predictions are relatively stable and what
level of theoretical uncertainty is present.
This paper concentrates on hybrid states, which are
formed by combining a gluonic
excitation with quarks.

Hybrids have been studied in the literature using
the flux tube model \cite{IP1,MP1,IP2,MP2,MPhD,IKP,CP},
the MIT bag model \cite{hybag},
an adiabatic heavy-quark bag model \cite{HHKR},
constituent gluon models \cite{HM,hycg},
and heavy-quark lattice gauge theory \cite{hylgt}.
In all these approaches
the lightest glueball
and hybrids
($H_q$, involving $u,d,s$ flavors) are predicted to have masses in the
$\approx 1{1\over 2}$-$2$ GeV region.
Hybrids are
very attractive experimentally
since they span complete flavor nonets and are
expected to include
the lightest $J^{PC}$-exotics
(which are forbidden to $q\bar q$).
For recent reviews of hybrids see \cite{hyrevs}.

Detailed predictions for hybrid spectroscopy
were first carried out using the MIT bag model and
QCD sum rules. The bag model predictions \cite{hybag}
suffer from parameter uncertainties
and possibly additional effects such as gluon self-energies,
so the absolute mass scale and the scale of
multiplet splittings are somewhat problematical.
Conclusions of the bag model
studies include the existence
of a lightest hybrid meson multiplet at $\sim 1.5$ GeV
and the presence of a $1^{-+}$ $J^{PC}$-exotic state in this multiplet.
In the bag model the lowest $q\bar q g$ hybrids have negative parity due to the
bag boundary conditions,
which give the first
TE gluon mode ($J^P=1^+$) lower energy than TM ($J^P=1^-$).
For heavy quarks it is unrealistic to assume a spherical
bag, so Hasenfratz, Horgan, Kuti and Richard\cite{HHKR} introduced
an adiabatic bag model in which the bag was allowed to deform in
the presence of fixed $Q\bar Q$ sources. The resulting $E(R)$ was used in
the two-body Schr\"odinger equation to give mass estimates for hybrids.
Masses found for the lightest hybrids were $\approx 3.9$ GeV
for $c\bar c$ (taken from their Fig.2) and 10.49 GeV for $b\bar b$.
The estimated systematic uncertainty for $b\bar b$ hybrids was $\pm 0.2$ GeV.

QCD sum rules have been applied to the study of hybrids, notably
the $1^{-+}$ and $0^{--}$ exotics,  by several
collaborations \cite{hysr1,hysr2,hysr3,hysr4,hysr5}.
Early results by these collaborations suggested a light
$1^{-+}$ exotic hybrid with a mass between
$\approx 1$~GeV and $\approx 1.7$~GeV.
The $0^{--}$ exotics were predicted to lie much higher,
at $3.1$-$3.65$~GeV.
Unfortunately, much of the
more recent work is not consistent with these results, although Balitsky,
Dyakonov and Yung (1986) continue to support a mass of $M(1^{-+})\sim$~1.5~GeV.
Latorre, Pascual
and Narison \cite{hysr2} cite higher masses of
$\approx 2.1$~GeV for the
$u,d$ $1^{-+}$
and $\approx 3.8$~GeV for the
$0^{--}$.
Govaerts {\it et al.}
\cite{hysr3}
estimate
$\approx 2.5$~GeV for
the $1^{-+}$
$q\bar qg$ ($q=u,d,s$), and their other exotic hybrid mass
estimates are rather higher than
previous references.
They
conclude however that the
sum rules for exotic hybrids are unstable, so all these results
are suspect.
For heavy
$1^{-+}$
hybrids Narison \cite{hysr2} estimates
4.1~GeV
for
$c\bar c$
and 10.6~GeV for
$b\bar b$.
In contrast, Govaerts {\it et al.}
find
$\approx 4.4$-$5.3$~GeV for
$c\bar c$ and
$\approx 10.6$-$11.2$~GeV for
$b\bar b$, {\it albeit} with reservations regarding the stability
of these results.
Thus, sum rules have reached no clear consensus regarding the masses of
hybrids, and recent results suggest rather higher masses than
previously thought.
Some technical errors in the earlier sum rule calculations have
been reported by Govaerts {\it et al.} \cite{hysr4}.
Sum rule calculations of decay couplings have also been reported;
deViron and Govaerts \cite{hysr5} anticipate a strong $\rho\pi$ decay mode for
the
$I=1$, $1^{-+}$ exotic.

Constituent gluon models for hybrids were introduced by Horn and Mandula
\cite{HM} and were subsequently developed by Tanimoto,
Iddir {\it et al.} and Ishida {\it et al.} \cite{hycg}.
Since these models assume a diagonal gluon angular momentum $\ell_g$ their
predictions for quantum numbers differ somewhat from the other models. For
the lightest hybrid states (with $\ell_g=0$) Horn and Mandula predict
nonexotic quantum numbers equivalent to $P$-wave $q\bar q$ states, since
the gluon has $J^P=1^-$. Exotic quantum numbers
including $1^{-+}$ are predicted in the
higher-lying $(\ell_{q\bar q}, \ell_g ) = (1,0)$ and $(0,1)$ multiplets.
Detailed spectroscopic predictions for hybrids
have not been published using
constituent gluon models, and the estimated masses are assigned large
uncertainties. A typical result, due to Ishida, Sawazaki, Oda and Yamada,
is 1.3-1.8 GeV for light nonexotic hybrids and 1.8-2.2 GeV for light exotics.
This type of model
predicts that the dominant two-body decay modes of light exotic
hybrids such as $1^{-+}$ are the $S+P$ combinations \cite{hycg}
such as
$b_1\pi$ and $a_1\pi$. This conclusion was subsequently supported by
studies
of
the flux tube model.

Lattice QCD will presumably give the most reliable
predictions for absolute hybrid masses,
although at present this
approach
has little to say about multiplet splittings.
In heavy quark lattice QCD,
in which
the $Q\bar{Q}$ pair is
fixed spatially and the gluonic degrees of freedom are allowed to
be excited,
the lightest charmonium hybrid was predicted by Michael {\it et al.}
\cite{hylgt} to have a mass of
$m({\it H}_c)_{quenched} = 4.04(3)\; {\rm GeV}$.
This reference adds an estimated shift of 0.15 GeV to compensate for the
quenched
approximation, which leads to a final lattice estimate of
$m({\it H}_c) = 4.19 \ {\rm GeV}$.
Note that a wide range of charm quark masses has been assumed in hybrid
spectrum calculations; in this HQLGT result a value of $m_c=1.32$ GeV was used,
whereas the flux tube calculations of Isgur, Merlin and Paton
\cite{MP1,IP2,MP2}
used $m_c= 1.77$~GeV. The sensitivity of the hybrid mass spectrum to
$m_c$ will be addressed subsequently.
The corresponding HQLGT estimates for $b\bar b $ hybrids were
$m({\it H}_b)_{quenched} = 10.56(3)\; {\rm GeV}$
and
$m({\it H}_b)   = 10.81 \ {\rm GeV}$.

In the flux tube model the more recent calculations
\cite{MP1,IP2,MP2} cite masses of about 1.9 GeV for the lightest
($q=u,d$) hybrid multiplet, about 4.3 GeV for $c\bar c$ hybrids
and about 10.8 GeV for $b\bar b$ hybrids. There is an overall variation
of about 0.2-0.3 GeV in these predictions, as indicated
in Table I. Although multiplet splittings are usually neglected in
the flux tube model,
a rather large inverted spin-orbit Thomas term
was found by Merlin and Paton \cite{MP2}.
The flux tube model also predicts very characteristic two-body decay modes
for hybrids \cite{IKP,CP} which have motivated
experimental studies of the channels $f_1\pi$
and $b_1\pi$, and suggest $h_1\pi$ and $\rho\pi$ \cite{CP}
as interesting future
possibilities.

The mass predictions for the lowest-lying $(1^{-+})$ exotic hybrid
(which is essentially the mass of the lightest hybrid multiplet) are
summarized in Table 1.
\eject

\begin{table}
\caption{Predicted $1^{-+}$ Hybrid Masses.}
\begin{tabular}{clll}
state & mass (GeV) & model & Ref. \\
\hline
$H_{u,d}$ &  1.3-1.8 & bag model & \cite{hybag} \\
          &  1.8-2.0 & flux tube model & \cite{IP1,MP1,IP2,MP2} \\
          &  2.1-2.5 & QCD sum rules (most after 1984) &
\cite{hysr2,hysr3,hysr4} \\
\hline
$H_c$     & $\approx 3.9$ & adiabatic bag model & \cite{HHKR} \\
          & 4.2-4.5 & flux tube model & \cite{MP1,IP2,MP2} \\
          & 4.1-5.3 & QCD sum rules (most after 1984) &
\cite{hysr2,hysr3,hysr4} \\
          & 4.19(3) \ $\pm$ sys. & HQLGT & \cite{hylgt} \\
\hline
$H_b$     & 10.49(20) & adiabatic bag model & \cite{HHKR} \\
          & 10.8-11.1 & flux tube model & \cite{MP1,IP2,MP2} \\
          & 10.6-11.2 & QCD sum rules (most after 1984) &
\cite{hysr2,hysr3,hysr4} \\
          & 10.81(3) \ $\pm$ sys. & HQLGT & \cite{hylgt} \\
\end{tabular}
\label{table1}
\end{table}

In this paper we carry out improved
numerical studies of the flux tube model, which is the
most widely cited model for hybrids. Previous flux tube estimates of the hybrid
spectrum made several simplifying assumptions, including a small oscillation
approximation and an adiabatic separation of quark and flux tube motion
\cite{IP1,MP1,IP2,MP2,MPhD}.
In principal
these could introduce important systematic biases in the spectrum.
We will present numerical results which are free of these approximations,
using a Hamiltonian Monte Carlo technique. Since our results for the lightest
hybrid
masses are quite similar to previous analytical results, we conclude that the
approximations made were reasonable, or when they did lead to important
numerical
inaccuracies (such as in the adiabatic approximation and in the small
oscillation
approximation at small R) the estimates of corrections to
the approximations were sufficiently accurate. Thus, we substantiate previous
estimates of hybrid masses in the flux tube model, and we also give masses for
higher
hybrid excitations using our techniques.

\section{The Flux Tube Model}

\subsection{Definitions}

In lattice QCD
widely separated static color sources are confined
by approximately cylindrical
regions of chaotic color fields\cite{FTLGT}.
The flux tube model is an attempt to describe this phenomenon with a simple
dynamical model,
and was motivated by the strong coupling expansion
of lattice QCD \cite{IP1} and
by early descriptions of
flux tubes as
cylindrical bags of colored fields \cite{GHKS}.
In this model one approximates the confining region between quarks by a string
of mass points, ``beads", with a confining potential between the beads.
Since a line
of flux in strong-coupling LGT
can be extended only in transverse directions (by the application of
plaquette operators), by analogy in the flux tube model
one allows only locally transverse spatial
fluctuations of the bead positions. For a
string of $N$ mass points which
connects a quark at site $0$ to an antiquark at site $N+1$ we write the
 flux tube model Hamiltonian as
\begin{equation}
H = H_{quarks} + H_{flux\ tube} \ ,
\end{equation}
\begin{equation}
H_{quarks} =
-{1\over 2 m_q} \vec \nabla_q^2
-{1\over 2 m_{\bar q}} \vec \nabla_{\bar q}^2
+ V_{q\bar q} \ ,
\end{equation}
\begin{equation}
H_{flux\ tube} =
-{1\over 2 m_b} \sum_{i=1}^N \bigg( \sum_{\hat \eta_T}
(\hat\eta_T\cdot\vec\nabla_i)^2 \bigg) +
\sum_{i=1}^{N+1} V\big( |\vec r_i - \vec r_{i-1}|\big) \ .
\end{equation}

Here $m_q$ and $m_{\bar q}$ are the quark and antiquark masses, $m_b$ is the
bead mass, and the $\{ \hat\eta_T\} $ are two orthogonal unit vectors
associated with
bead $i$ that are transverse to the local string tangent $(\vec
r_{i+1}-\vec r_{i-1})/|\vec r_{i+1}-\vec r_{i-1}|$.
In this study we use a standard
linear form for the string potential,
\begin{equation}
V\big(|\vec r_i - \vec r_{i-1}|\big) = a|\vec r_i - \vec r_{i-1}| \ ,
\end{equation}
and we usually set the string tension $a$ equal to $1.0$ GeV/fm.
For our estimates of physical hybrid masses we will augment this with
a color Coulomb interaction for $V_{q\bar q}$ in (2).

\subsection{Adiabatic Potentials and Flux Tube Parameters}

In the flux tube studies of Isgur, Kokoski, Merlin, and Paton
\cite{IP1,MP1,IP2,MP2,MPhD}
the combined quark and flux tube system is treated using an adiabatic
approach as a zeroth order approximation.
In the adiabatic analysis one exploits the anticipated
fast dynamical response of the flux tube
relative to heavy-quark time scales, and separates
the flux tube and quark degrees of freedom.
This is accomplished by
fixing the $q\bar q$ separation at $R$
and
determining an eigenenergy $E_{\Lambda}(R)$ of the flux tube.
Solution of the Schr\"odinger equation
for the $q\bar q$ wavefunction in the flux tube ground state
potential $E_0(R)$ then gives the conventional
$q\bar q$ meson spectrum in the adiabatic approximation.
Hybrids are excited states of the string in this
approach, and are found using an excited string potential $E_{\Lambda}(R)$.
The
lightest hybrid follows from an $E_1(R)$ in which the lowest string mode has a
single orbital excitation about the $q\bar q$ axis.

In previous studies the adiabatic potentials $\{ E_{\Lambda}(R) \} $ were
determined {\it assuming small string fluctuations relative to
the $q\bar q$ axis}.
We shall find that this is an inaccurate approximation for typical hadrons,
assuming $R\approx 1$ fm.

One motivation for the small oscillation approximation is that it
leads to relatively simple analytical results; when applied to (3)
it gives a quadratic Hamiltonian,
which can be diagonalized using Fourier modes.
To illustrate this, consider a string with fixed ends at ${\bf x}_0 = (0,0,0)$
and ${\bf x}_{N+1} = (0,0,R)$ and
$N$ dynamical beads, with motion allowed only in the transverse $\{ x_i,y_i\}$
directions.
In the small oscillation approximation, assuming that the beads
are equally spaced in $z$ by $a_0$, so $z_n = n a_0$ and $a_0 = R/(N+1)$,
the flux tube Hamiltonian becomes
\begin{equation}
H_{flux\ tube} = aR
-{1\over 2m_b} \sum_{i=1}^N
\bigg( {\partial^2\over \partial x_i^2}+
{\partial^2\over \partial y_i^2}\bigg)
+{a/a_0\over 2} \sum_{i=1}^{N+1}
\bigg((x_i-x_{i-1})^2+
(y_i-y_{i-1})^2\bigg)
\ .
\end{equation}
This is equivalent to a system of $N$ coupled masses $\{ m_b\} $ with an
effective
spring constant of $k = a/a_0 =  (N+1) a / R$.
We can diagonalize this using sine variables
\begin{equation}
s_{n,\lambda=(1,2)} = \sqrt{2\over N+1} \ \sum_{i=1}^N \; \sin (k_n z_i ) \,
(x,y)_i
\end{equation}
and
\begin{equation}
(x,y)_i = \sqrt{2\over N+1} \ \sum_{n=1}^N \; \sin (k_n z_i ) \,
s_{n,\lambda=(1,2)}
\end{equation}
where $k_n = \pi n / R$.
This gives
\begin{equation}
H_{flux\ tube} = aR
+ \sum_{n=1}^N\sum_{\lambda=1}^2
\bigg(
-{1\over 2m_b}
{\partial^2\over \partial s_{n\lambda}^2}
+{1\over 2}
\, \kappa_n\,
s_{n\lambda}^2 \bigg)
\end{equation}
where the effective spring constant of the $n$th Fourier mode is
\begin{equation}
\kappa_n = {4 (N+1) a \over R} \; \sin^2 \bigg( {\pi n \over 2 (N+1) } \bigg)
\ .
\end{equation}

The ground state
energy of the string, which is used as the adiabatic potential
for conventional ($q\bar q$) mesons,
is $aR$ plus the sum of $\omega/2$ for each mode
in the small oscillation
approximation.
The individual eigenfrequencies are
\begin{equation}
\omega_n = \sqrt{ \kappa_n / m_b} =  2\; \sqrt{ { (N+1) a \over m_b R } } \;
\sin \bigg( {\pi n\over 2 (N+1)} \bigg) \ ,
\end{equation}
and
the mode sum runs over $n=1$ to $N$ and $\lambda=1,2$.
The resulting ground state energy is
\begin{equation}
E_0(R) =
aR + \sum_{modes} {1\over 2}\, \omega_{n}
= aR +  \sqrt{ {2(N+1) a\over m_b R } }
\
\Bigg\{
{ \sin \Big( {\pi N\over  4 (N+1)} \Big)
\over
\sin \Big( {\pi \over  4 (N+1)} \Big) }
\Bigg\}
\ ,
\end{equation}
which agrees with the result of Isgur and Paton \cite{IP1}.
The most general adiabatic potential in the small oscillation approximation is
\begin{equation}
E(R) = E_0(R) +  \sum_{modes\atop  m} n_m \; \omega_m(R) \ ,
\end{equation}
where $n_m$ is the number of excitations of the $m$th flux tube
mode.

The ground state wavefunction of the string in the small oscillation
approximation is a Gaussian in the Fourier mode amplitudes,
\begin{equation}
\Psi_0(\{ x_i, y_i \} ) = \prod_{n,\lambda }
\eta_n \, e^{ -  s_{n\lambda}^2/2\sigma_n^2 } \ ,
\end{equation}
where the Gaussian width of mode $n,\lambda$ is given by
\begin{equation}
\sigma_n =
{1\over \sqrt{ m_b\omega_n}} =
{
\bigg[ {R \over (N+1) a m_b }\bigg]^{1/4}
\over
\bigg[ 2\sin\Big({\pi n \over  2 (N+1) }\Big) \bigg]^{1/2}
}
\ .
\end{equation}
This suggests an estimate of the range of validity of the small oscillation
approximation; it
should fail
when these fluctuations become comparable to $R$.

Excitations can be created from the ground state wavefunction (13)
through the application of ``phonon" creation operators
\begin{equation}
A^{\dagger}_{n,\lambda} = {1\over \sqrt{2 m_b \omega_n }}\;
\bigg( -{\partial \over \partial s_{n\lambda}
}+ m_b \omega_n  s_{n\lambda} \bigg) \ ,
\end{equation}
with an increase in energy of $\omega_n$. States with definite angular
momentum component $\Lambda$
along the $q\bar q$-axis, which are useful in constructing
hybrid states, are created by the linear combinations
\begin{equation}
A^{\dagger}_{n,\Lambda=\pm 1} = {1 \over \sqrt{2}}\bigg(
\mp A^{\dagger}_{n,1} - i
A^{\dagger}_{n,2}  \bigg) \ .
\end{equation}

The flux tube parameters $a, m_b$ and $N$ can be constrained by the plausible
requirement that the maximum propagation velocity on the flux tube be $c$. In
the large-$N$ limit this implies (from (10))
\begin{equation}
v_{max}/c \equiv \lim_{k\to 0} \ {\partial \omega \over \partial k } =
\sqrt{ { a a_0 \over m_b  } } = 1 \ .
\end{equation}
The length $a_0$ might
reasonably be identified with the transverse flux tube extent of
$\approx$~$0.2$-$0.3$~fm
found in a lattice Hamiltonian string theory \cite{MP3}
or the
$\approx$~$0.2$-$0.4$~fm estimated in
lattice
Monte Carlo QCD \cite{FTLGT}.
For a typical string tension of $a = 1.0$
GeV/fm the constraint (17)
implies $m_b\approx 0.2$-$0.4$ GeV.
We take $m_b=0.2$ GeV as our standard value, since the larger transverse extent
of 0.4 fm may
represent fluctuations of an intrinsically smaller flux tube.

Isgur, Merlin and Paton \cite{IP1,MP1,IP2,MP2}
also treat $a_0$ as a fundamental length but allow $N$ to vary
continuously with $R$, so that $a_0=R/(N+1)$ is constant.
The large-$R$ hybrid potential gap of
\begin{equation}
\lim_{R\to \infty} \omega_1(R) =
\sqrt{ a  \over m_b }\;  {\pi \over \sqrt{ (N+1) R }}
\end{equation}
then becomes
\begin{equation}
\lim_{R\to \infty} \omega_1(R) =
\sqrt{ a a_0 \over m_b }\; \cdot {\pi \over R } =
{\pi \over R } \ .
\end{equation}
The
final result follows from the constraint (17). An excitation
energy of $\pi / R$ was found earlier
by Gn\"adig {\it et al.} \cite{GHKS}
in their cylindrical bag
model of a flux tube.

Of course
we cannot vary $N$ continuously in a numerical
simulation.
In this first numerical study we
shall mainly consider the simplest fixed-$N$ case, $N=1$.
As we shall see, this allows a detailed study of the
various approximations used previously in estimating hybrid masses, and
leads to very plausible results for conventional and hybrid spectroscopy.

\section{Numerical Results for Adiabatic Potentials}

We will now generate adiabatic potentials numerically, for
comparison with
the small oscillation potentials
derived in the previous section.

The adiabatic $N=1$ (single bead) problem can be integrated numerically,
since there is
only motion in a single plane, and the bead wavefunction can be separated as
$\Psi_{\Lambda}(\rho,\theta) = \psi_{\Lambda}(\rho) \exp ( i \Lambda
\theta)$. The ordinary differential equation satisfied by
$\psi_{\Lambda}(\rho)$ is
\begin{equation}
-{1\over 2 m_b} \bigg(
{d^2 \psi_{\Lambda} \over d \rho^2} +
{1\over \rho} {d \psi_{\Lambda} \over d \rho} \bigg)
+ \bigg( 2 a \sqrt{\rho^2 + R^2/4 } + {\Lambda^2 \over 2 m_b \rho^2 }
\bigg) \psi_{\Lambda}  = E_{\Lambda}(R) \psi_{\Lambda} \ ,
\end{equation}
and the exact $q\bar q$ meson adiabatic potential $E_0(R)$ and first hybrid
adiabatic
potential
$E_1(R)$ follow from solving this equation for its lowest eigenvalue with
$\Lambda=0$ and $\Lambda=1$ respectively. The potentials
$E_0(R)$ and $E_1(R)$ and the potential gap $E_1(R) - E_0(R)$
are shown in Figs.1 and 2 for $m_b=0.2$ GeV and $a=1.0$ GeV/fm.
In the limit of infinitely massive quarks the adiabatic approximation
is exact, the $Q\bar Q$
separation approaches zero, and the hybrid mass gap
is therefore
$E_1(0)-E_0(0)$ ($=0.829$ GeV with these parameters).
As $R$ increases the potential gap falls, but
asymptotically as $2\sqrt{a/ m_b R}\ $ ((10) with $n=1$ and $N=1$)
rather than as the $\pi / R$ of Isgur
and Paton, due to our assumption of a fixed-$N$ flux tube.
The small oscillation
adiabatic potentials and gap
from (10-12) are shown as dashed lines in Figs.1 and 2; they are evidently
useful
only beyond $R\approx 1$ fm.
Since $R\approx 1$ fm is a typical light $(u,d,s)$ hadron length scale, the
small
oscillation approximation is inappropriate
for light hadrons.
For smaller $R$ the approximate small oscillation adiabatic potentials
depart considerably from the true $\{ E_{\Lambda}(R) \}$ (solid lines),
and actually diverge as $R\to 0$.

In the previous section we suggested a condition for applicability of the
small oscillation approximation, which is that $R$ should be
much larger than the zero-point fluctuations $\sigma_n$ in the string ground
state. The largest fluctuations are in the $n=1$ mode; taking this case,
the mode width for $N=1$ is
\begin{equation}
\sigma_1 = \bigg[ {R \over 4 m_b a} \bigg]^{1/4} \ .
\end{equation}
Note the weak parameter dependence of the scale of fluctuations implied
by the $1/4$ power. The characteristic length $R_c$ at which
the scale of fluctuations $\sigma_1$
equals $R$ is given by
\begin{equation}
R_c(N=1) =   (4 m_b a )^{-1/3} = 0.37\ {\rm fm}.
\end{equation}
$R$ should be significantly larger than this for the small
oscillation approximation to be useful, which is supported by our Figs.1 and 2.

Although this paper is primarily concerned with numerical results for the $N=1$
one-bead flux tube model, we can carry out simulations for larger $N$ using
a Hamiltonian Monte Carlo technique\cite{GRW}. This method will be discussed
in the next section, in which it is applied to
the combined dynamical quark and flux-tube system. As a test of the Monte Carlo
method
we confirmed that the adiabatic potentials $E_0(R)$ and $E_1(R)$ with $N=1$ are
accurately reproduced (Fig.2),
and we also show results for the $N=2$ case.
The hybrid mass gap apparently falls rapidly
with increasing $N$, so it may be difficult to find
a realistic description of the spectrum with a fixed-$N$ flux tube model for
larger $N$; the excitation energy of a many-bead string is presumably quite low
relative to the
$N=1$ case, assuming similar $m_b$ and $a$.
There are also rather subtle complications in the dynamics of the $N>1$ flux
tube
with fixed ends\cite{bswan};
the constraint of transverse bead motion implies dependence of
energies on the initial conditions,
which must then be varied to find the lowest-lying
state.

\section{Hybrids with Dynamical Quarks}

\subsection{Adiabatic Results}

Thus far we have only considered the adiabatic potentials. Now we shall solve
the
two-body $q\bar q$ Schr\"odinger equation
in the exact adiabatic potentials $\{ E_{\Lambda}(R) \} $,
which are determined by numerically integrating (20) for a flux tube with
static
sources separated by $R$. The flux tube ground state and first excited state
potentials $E_0(R)$ and $E_1(R)$ lead to conventional and the lightest hybrid
mesons
respectively.

For hybrids there is a centrifugal barrier for the $q\bar q$
pair that arises from the matrix element of $\vec L_q^2$ in the full
quark-and-flux-tube angular momentum eigenstate. The angular wavefunction of
the combined gluon or flux tube and quark system was
discussed by Horn and Mandula
\cite{HM} and subsequently by
Hasenfratz {\it et al.} \cite{HHKR} and Isgur and Paton \cite{IP1}.
There are discrepancies
between these references in the $C$ and $P$ hybrid quantum numbers; this does
not affect our conclusions regarding hybrid energies
because of degeneracies between the levels
concerned.
The latter two references give essentially the same rigid body
angular wavefunction
for the full system, which is
\begin{equation}
\psi_H^{(L)}  \propto  \ {\cal D}^{\; (L)}_{M \Lambda}(\phi, \theta, -\phi) \ .
\end{equation}
(The Hasenfratz {\it et al.}
wavefunction does not have the final $-\phi$ argument because it uses
body-fixed rather than space-fixed coordinates.)
This is the amplitude to find the $q\bar q$ axis pointing along
$(\theta, \phi)$
in a hybrid state with total orbital angular momentum $L$ and
$\hat z$-projection $M$, and $\Lambda$ is the projection of the flux tube
orbital
angular momentum along the $q\bar q$ axis.
$\Lambda = n_{m+}- n_{m-}$, where $n_{m\pm}$ is the number of excitations
of the m$th$ flux tube mode, ($+$) for right-handed and ($-$)
for left-handed, as in (16).
Thus for a single flux tube excitation $\Lambda = \pm 1$, for
doubly-excited flux tubes $\Lambda = 0, \pm 2$, and so forth.
Parity implies a degeneracy between
$\Lambda = \pm |\Lambda|$ levels, so without loss of generality
we assume nonnegative $\Lambda$ in
our simulations.
The total orbital angular momentum $L$ is constrained
to be $L\geq |\Lambda|$.

The wavefunction (23) is not fully diagonal in configuration space;
it assumes that the flux tube is in a coherent superposition of orientations
about
the $q\bar q$-axis such that the angular
momentum
projection $\Lambda$ along the $q\bar q$ axis is diagonal. This requires a
wavefunction
\begin{equation}
\psi_{f.t.}^{(\Lambda)}(\phi_b) = {1\over \sqrt{2\pi}} \; e^{i\Lambda \phi_b} \
,
\end{equation}
where $\phi_b$ gives the rotation of the flux tube about the $q\bar q$-axis
relative to a
reference configuration.
In our Monte Carlo we used basis states which are fully diagonal in coordinate
space,
so a configuration is defined (for $N=1)$
by the coordinates $\vec x_q, \vec x_{\bar q}, \vec x_b$, which implicitly
determine its orientation relative to
a reference configuration and space fixed axes, specified by the $q\bar q$-axis
angles
$\theta,\phi$ and the rigid body rotation angle $\phi_b$.
This relation is defined by the effect of the rotation
operator,
\begin{equation}
|\theta,\phi,\phi_b\rangle = e^{-i\phi J_z}\; e^{-i\theta J_y}\;e^{+i\phi
J_z}\;
|\hat z,\phi_b\rangle \ .
\end{equation}
The angles $\theta$ and $\phi$ are specified trivially by the $q\bar q$ axis.
The rigid body
rotation angle $\phi_b$ is rather more complicated, and satisfies
\begin{equation}
\sin ( \phi_b) =
{
\sin (\phi )
(x_b - x_{q\bar q\ cog} )
+
\cos (\phi )
(y_b - y_{q\bar q\ cog} )
\over
|\vec r_b - \vec r_{q\bar q \ cog} |
} \ ,
\end{equation}
as may be confirmed from Fig.3, which shows the operations required to
reach a general
configuration from an unrotated ``reference" configuration.

Given the $\phi_b$ dependence implicit in the $\Lambda$ states,
our $\phi_b$-diagonal angular wavefunctions must
be of the form
\begin{equation}
\langle \theta,\phi,\phi_b| L,M\Lambda\rangle  \propto
\ {\cal D}^{\; (L)}_{M \Lambda}(\phi, \theta, \phi_b -\phi) \ ,
\end{equation}
which we shall use as the guiding wavefunction for hybrid states in the Monte
Carlo
simulation.

In their equation (28) Isgur and Paton \cite{IP1} (see also equation (6)
of Merlin and Paton \cite{MP1})
introduce a simple approximation for the
matrix element of $\vec L_q^2$,
which neglects a mixing operator that
raises and lowers $\Lambda$. This approximation gives
$\langle\vec L_q^2\rangle \approx L(L+1) - \Lambda^2$, which transforms the
Schr\"odinger equation into an ordinary
differential equation for the adiabatic $q\bar q$ radial wavefunction
$\psi_{\Lambda}^{(L)}(r)$,
\begin{equation}
H_{adia.} = -{1\over 2 \mu}
\bigg({\partial^2\over \partial r^2} +
{2\over r} {\partial\over \partial r} \bigg) +
{L(L+1) - \Lambda^2 \over 2\mu r^2} + E_{\Lambda}(r) \ ,
\end{equation}
\vskip 0.5cm
\begin{equation}
H_{adia.}
\psi_{\Lambda}^{(L)}(r) = M_H
\psi_{\Lambda}^{(L)}(r)  \ .
\end{equation}
Isgur and Paton determined the hybrid spectrum by solving this eigenvalue
problem,
with an additional approximation; they replaced the singular
small oscillation adiabatic
potentials $E_{\Lambda}(R)$ (12) with approximate forms that were nonsingular
at $R=0$.
We shall instead use the exact (numerical) adiabatic potentials
$\{ E_\Lambda (R) \} $ (from (20))
in (28,29) above, which gives the true adiabatic result
for the spectrum. This will be compared to our Monte Carlo results.

\subsection{Monte Carlo Simulation}

We improve on previous studies of the flux tube model by using
the Guided Random Walk (GRW) Hamiltonian Monte Carlo
algorithm \cite{GRW}
to solve the full $N=1$ model without adiabatic or small oscillation
approximations.
The GRW algorithm maps the imaginary time Schr\"odinger
equation onto a diffusion
problem, which is then solved
numerically using weighted random walks in the configuration
space of the system. The statistical error is reduced through the use of
a guiding wavefunction for importance sampling, which is used to determine
stepping probabilities between configurations during the walk. This importance
sampling does not bias
the energies and matrix elements.

In this algorithm a random walk is generated by stepping in the coordinates
which define
configuration space.
For a $q,\bar q$ and $N$-bead system
there are $N_x=2N+6$ possible coordinates to increment.
Starting from a specified initial
configuration of quark, antiquark and bead locations at $\tau=0$, one of the
coordinates is chosen at random, and an increment
$x\to x + h_q $ (or $h_b$)
is made in that coordinate with probability
\begin{equation}
P(step) = {1\over 2} {\psi_g(x_{new}) \over \psi_g(x_{current}) } \ .
\end{equation}
If the move is not accepted, a move in the opposite direction is made,
$x\to x - h_q $ (or $h_b$).
The step sizes in
$h_b$ (for bead moves)
and
$h_q$ (for quark or antiquark moves, with $m_q$ and $m_{\bar q}$
assumed equal)
are given by
\begin{equation}
h_b =  \sqrt{ { N_x h_\tau \over m_b} }
\end{equation}
and
\begin{equation}
h_q =  \sqrt{ { m_b \over m_q} } \; h_b \ ,
\end{equation}
where $h_\tau$ is a small step size
in Euclidean time (relative to inverse energy scales).
After each move the Euclidean time
is incremented by $h_\tau$.
Excited states with nodes in the guiding wavefunction $\psi_g$
require
special consideration; for these cases
we test
that moves do not cross the nodal surface, and if they do they are
rejected and another move is generated. This introduces a bias which
vanishes as $h_\tau \to 0$.
There is also a bias in excited states if a guiding wavefunction is used which
has
incorrect nodes.

For the
static quark simulations in Sec.III we used a guiding wavefunction which is
a Gaussian in the total string length $R_{str}$,
\begin{equation}
\psi_g = \exp\bigg\{ - (R_{str}/\xi )^2 \bigg\} \ ,
\end{equation}
and allowed only bead moves. The optimum guidance parameter $\xi$ was estimated
numerically by minimization of the statistical error, specifically by
minimizing the
variance of the weight factor $w(\tau)$ in (35).
For $N=1$ and all the $R$ values considered here
the optimum value was found to be $\xi\approx 1.5$ fm.

For the dynamical quark ground state we use as our guiding wavefunction
\begin{equation}
\psi_g^{(0)} = \exp\bigg\{ - (R_{str}/\xi )^2 -
R/ \xi_{q\bar q} \bigg\} \ .
\end{equation}
This simple generalization of
the static quark Gaussian (33) includes
a suppression of the wavefunction with increasing
interquark separation
$R$ for fixed string length
$R_{str}$,
as is intuitively
expected for heavy quarks.
For excited-$L$ $q\bar q$ and hybrid states the wavefunction is more
complicated, and must incorporate nodes to ensure orthogonality to the
ground state (see below).

In the course of a random walk
from Euclidean time $0$ to $\tau$ we generate a path-dependent
weight factor,
given by
\begin{equation}
w(\tau) = \exp
\Big\{  \int_0^\tau \bigg(
- V
+
\bigg[
{\nabla_q^2 \psi_g + \nabla_{\bar q}^2 \psi_g  \over 2 m_q }
+  {\nabla_b^2 \psi_g \over 2 m_b } \bigg]\, \psi_g^{-1} \; \bigg)
d\tau \; \Big\} \ ,
\end{equation}
where the Laplacians are in the $6$ quark and antiquark and $2N$
(transverse) bead coordinates
respectively.  The form (35) and the step sizes $h_b$ and $h_q$ above are
chosen so
that a histogram of these weights in configuration space $\{ x \} $
is proportional to a solution $\psi(\{ x \} , \tau ) $
of the Euclidean time Schr\"odinger equation. Actually $w(\tau)$ gives the
related function
$\psi_g(\{ x\} )\psi(\{ x\} ,\tau))$ \cite{BK};
this $\psi_g \psi$
can also be used to determine the
ground state energy,
and is generated with a smaller
statistical error than $\psi$ itself.
The energy is determined from the large-$\tau$ behavior of the weight $w(\tau
)$:
At
large $\tau$
the walk-averaged weight $<w(\tau )>$
approaches an exponential in $\tau$,

\begin{equation}
\lim_{\tau\to\infty} <w(\tau )>  = \kappa \,
e^{-E_0\tau}\bigg(1 +
O(e^{-E_{gap}\tau}) \bigg)  \ .
\end{equation}
so we may determine $E_0$ from measurements of
$<w>$ at two successive Euclidean times,
\begin{equation}
E_0 = \lim_{\tau_1,\tau_2\to\infty} {1\over (\tau_2 - \tau_1 )}
\ln \Bigg\{ {<w(\tau_1 )> \over <w(\tau_2 )> } \Bigg\} \ .
\end{equation}
In practice we leave $\tau_2-\tau_1$ fixed and increase $\tau_1$ until
the $E_0$ estimate has converged to the required accuracy.

If a guiding wavefunction $\psi_g$ with nodes is used, we recover
the lowest energy
eigenvalue for which $\psi=0$ on those nodes. If the nodes are identical to
those of an excited
state $\psi_n$ of the system, we recover the correct $E_n$ from (37).

This algorithm gives the true eigenenergy for any
guiding wavefunction
$\psi_g$ with correct nodes,
provided that the initial configuration
has nonzero amplitude in the ground state. The results become statistically
more accurate as the guiding wavefunction is made closer to the
true eigenfunction $\psi_n$, and one may confirm that the
best possible choice is an energy eigenfunction, $\psi_g = \psi_n$ \cite{BK}.
In this case the weight factor (35)
becomes $w=exp(-E_n\tau)$ exactly for each walk, so the
energy can be determined from a single walk at arbitrary $\tau$.
Of course we do not know $\psi_n$ in general, so we use a parametrized
Ansatz for $\psi_n$ as our $\psi_g$, and determine the optimum
parameters numerically by minimizing the variance of the weight
factors $\{ w \}$ in a sample of random walks.
Given the optimized guiding wavefunction $\psi_g$, we then determine
$E_n$ using (37).

\subsection{Monte Carlo Results}

For
$\alpha_s=0$ we generated Monte Carlo energies for quark masses of
$m_q= 0.33$,
0.5, 1.0, 1.5, 2.5, 5.0 and 10.0 GeV, with a string tension of $a=1.0$ GeV/fm.
The optimized guiding wavefunction parameters in (34) were $\xi = 1.5$ fm
and $\xi_{q\bar q} = 1.4$, 1.0, 0.7, 0.6, 0.5, 0.4  and 0.3 fm
for the quark masses given above.
The Euclidean
times used, which were chosen to insure
convergence to ground state results to within our statistical errors, were
$\tau_1 = 10.0$ GeV$^{-1}$ and
$\tau_2 = \tau_1 + 1.0$ GeV$^{-1}$, and the step size was $h_\tau = 0.005$
GeV$^{-1}$. For energy differences of
excited and ground state levels, $E_n-E_0$, we found
adequate convergence with a smaller time of $\tau_1=5.0$ GeV$^{-1}$.
We also generated energies for various
other guidance and time parameters to confirm the accuracy of these results.
The sample size was usually $N_{rw}=8\times 1024$ walks
(8 separate runs to generate errors), and we used bootstrap on each of the
8 runs
to suppress dependence
on the initial configuration.
(In a bootstrapped run the final configuration
of a walk at $\tau = \tau_2$ is used as the initial configuration of
the next walk at $\tau=0$.)
For hybrids with $m_q=0.33$ and 0.5~GeV
we used longer runs
of $N_{rw} = 8\times 4096$ walks to compensate for the larger statistical
errors.

The adiabatic ground state energies (from (28,29) with the
potential $E_0(R)$ of (20))
and Monte Carlo results for $N=1$
are summarized in Table 2
for
$\alpha_s=0$, $m_b=0.2$ GeV and $a=1.0$ GeV/fm.

\begin{table}
\caption{Adiabatic and Exact (Monte Carlo) Ground State Energies for
$N=1$. }
\begin{tabular}{ddd}
$m_q$ (GeV) & $E_0^{\rm adiabatic}$ (GeV) & $ E_0^{\rm Monte Carlo} -
E_0^{\rm adiabatic}$  (GeV) \\
\hline
 0.33 & 1.985 & 0.274(4) \\
 0.50 & 1.868 & 0.231(5) \\
 1.00 & 1.711 & 0.187(3) \\
 1.50 & 1.638 & 0.164(3) \\
 2.50 & 1.563 & 0.148(3) \\
 5.00 & 1.484 & 0.124(2) \\
 10.0 & 1.425 & 0.114(3) \\
\end{tabular}
\label{table2}
\end{table}

Evidently the adiabatic approximation considerably underestimates the
ground state energy, by up to 0.3 GeV for light $(u,d)$
quark systems. The
discrepancy falls rather
slowly with increasing quark mass, approximately
as $m_q^{-1/4}$.

For excited-$L$ quarkonia we generalize the ground state
guiding wavefunction to
\begin{equation}
\psi_g^{(L)} =
\psi_g^{(0)} \cdot R^{\, L} \cdot f(\theta, \phi) \ ,
\end{equation}
where the angular function depends on the direction of the $q\bar q$
axis, and was taken to be the real part of
$Y_{LM}(\theta, \phi)$.
(The algorithm requires a real wavefunction for importance sampling.)
The
radial factor $R^{\, L}$ is not essential but is expected to be closer
to the true $\psi_0^{(L)}$, and its inclusion reduces our statistical errors
somewhat.

For hybrid states
the amplitude to find the system at
$(\theta, \phi, \phi_b)$
is given by (27)
\begin{equation}
\psi_H^{(L)}
(\theta, \phi, \phi_b)
\propto  \ {\cal D}^{\; (L)}_{M \Lambda}(\phi, \theta,
\phi_b - \phi ) =
e^{i\Lambda \phi_b} \
e^{i(M-\Lambda) \phi} \ d^{\; (L)}_{M\Lambda}(\theta)
\ .
\end{equation}
For our full hybrid guiding wavefunction we multiply the real part of
this angular function by
a radial wavefunction similar to our ground state $\psi_g$,
\begin{equation}
\psi_g^{(H)} =
\psi_g^{(0)} \cdot  \rho_b \, R \cdot
f(\theta, \phi,\phi_b )
\ ,
\end{equation}
\begin{equation}
f(\theta, \phi, \phi_b)
=
\; d^{\; (L)}_{M\Lambda}(\theta)
\; \cos (\Lambda \phi_b +
(M-\Lambda) \phi )
\ .
\end{equation}
The product of $\rho_b$ (the bead-axis distance)
and $R$ was introduced as a simple centrifugal suppression factor.

There is a systematic bias in our results for excited states due to the
nodal surfaces specified by the angular wavefunctions $f$; these
surfaces
are exact only in the limit $m_q\to\infty$.
For our high statistics quarkonium simulations
we used $M=0$ states for simplicity, since they are $\phi$-independent.
We checked for evidence of node bias
by comparing the energies found using guiding wavefunctions
with different
magnetic quantum number $M$, which have different nodal surfaces.
The bias in $q\bar q$ states was at most about 10 MeV, comparable to our
statistical errors.
For the ${}_1P$ hybrid however we found a significant $M$-dependent
bias; in Fig.4 we show hybrid energies determined using both $M=0$ and
$M=1$ in (41). The largest bias was at the smallest quark mass of $m_q=0.33$
GeV, for which we found $E({}_1P,M=1) - E({}_1P, M=0) = 52(18)$ MeV.
This bias will be discussed in more detail in our treatment of hybrids
with physical parameters.

Fig.4 shows
the $P$-wave and $D$-wave $q\bar q$ levels and the first
hybrid level (${}_\Lambda L = {}_1P$) relative to the
ground state energy $E_0$, using both the adiabatic approximation (lines)
and Monte Carlo (points).
Our results show that the
adiabatic approximation is more accurate
for the
energy differences $\{ E_n - E_0\} $,
which are the experimentally observable quantities,
than
for $E_0$ itself.
The
largest discrepancies
between adiabatic and Monte Carlo results
are $\approx 100$~MeV, for the $D$-wave and hybrid
states at the lightest quark mass of $0.33$ GeV.  Note that the
adiabatic approximation {\it overestimates} the excited-$L$ energies but
{\it underestimates} the hybrid energy. Thus, {\it
if we use the adiabatic approximation
and fit the
experimental $D$-wave levels, we
underestimate the light hybrid mass by $\approx$ 200 MeV}.

In their
analytical study of the flux tube model, Merlin and Paton
\cite{MP1} also found that postadiabatic corrections
reduce the excited-$L$ energies and increase
the hybrid energy. They find $(q=u,d)$ $P,D$ and
${}_1P$ hybrid energy shifts
which are quite similar in relative strength to our Monte Carlo results;
this led Isgur and Paton to revise their adiabatic hybrid mass estimate
upwards from 1.67~GeV to $\sim $1.9~GeV \cite{IP2}.
The overall scale of adiabatic corrections quoted by Merlin and Paton
\cite{MP1} (see especially their Table 6)
is about twice as large as we find numerically, but
this may be due to their use of the large-$N$ limit,
whereas we have specialized
to $N=1$.

\subsection{Physical Hybrid Masses}

The flux tube results discussed in the previous section are not applicable to
real hadrons because they do not include the attractive color
Coulomb interaction. Without the Coulomb interaction the flux tube at small
$R$
gives an SHO-like adiabatic potential (see $E_0(R)$ in Fig.1), which leads
to nearly equal $S$-$P$-$D$
splittings in the spectrum of conventional $q\bar q$
mesons (as in Fig.4). A realistic description of the
$S$-$P$-$D$ splittings requires the familiar ``funnel shaped" potential,
in which linear confinement is augmented by a short ranged attraction.

In conventional potential models the Coulomb plus linear form
\begin{equation}
V_{q\bar q}(R) =
-{4\over 3} {\alpha_s\over R} + aR + V_0
\end{equation}
is most often used, with a string tension of $a\approx 0.9$-$1.0$
GeV/fm giving
the best fit. Perturbative QCD predicts that the
effective Coulomb interaction strength $\alpha_s$ should
run with the scale of momentum of the scattered constituents,
provided that we are
well above any intrinsic mass scales. For resonance physics this
requirement is obviously
not satisfied, but there is nonetheless clear evidence for
a rapid decrease of $\alpha_s$ with increasing quark mass;
fits to spectroscopy typically require
$\alpha_s\approx 0.6$-$0.7$
for
$q=u,d,s$,
$\alpha_s\approx 0.3$-$0.4$
for $q=c$
and
$\alpha_s\approx 0.2$
for $q=b$.

For our realistic parameter set we assume constituent quark masses of
$m_q=0.33, 0.55$ and $1.5$~GeV for $q=u(d),s$ and $c$, and
again set the string tension equal to
$a=1.0$~GeV/fm.  In addition we include
a color Coulomb and constant potential,
\begin{equation}
V_{q\bar q} =
-{4\over 3} {\alpha_s^{ft}\over R} + V_0
\end{equation}
in the flux tube quark Hamiltonian (2). The additive constant $V_0$
is found to be large and negative in potential models,
and in the flux tube model is required in part to cancel the zero-point
energies of the
beads. The coefficient $-4/3$ multiplying $\alpha_s / r$ in the color
Coulomb interaction merits additional comment.
In constituent gluon models of hybrids the $q\bar q$ pair would be in a color
octet,
so the $-4/3$ would be replaced by $1/6$. In the flux tube model, in which
gluonic excitations are presumed nonperturbative in $\alpha_s$, it may be more
realistic to
use $-4/3$. This can be motivated by noting that at small $R$ the
lowest gluonic excitation is a color singlet $q\bar q$ pair (hence $-4/3$)
plus a scalar glueball, rather
than a $q\bar q$ color octet pair with a diverging $+1/6$ color Coulomb
interaction
\cite{NIft}.

The $\alpha_s^{ft}$ in the $N=1$ flux tube $V_{q\bar q}$
cannot be
compared directly to the Coulomb plus linear $\alpha_s$, because the
fixed-$N$ flux tube gives an SHO-like
confining potential at
short distances
(see $E_0(R)$ in Fig.1)
in addition to the linear term which dominates at large $R$.
Since $\alpha_s^{ft}$ in the fixed-$N$ flux tube model must
cancel this additional contribution
to produce a
funnel shaped potential comparable to the standard Coulomb plus linear form,
it is
larger than the
potential model $\alpha_s$.

We used
multiplet-averaged $E_S$ and $E_P$ energies as input to fix
$\alpha_s^{ft}$ and $V_0$
in each flavor sector.
The numbers used
were
$E_P-E_S = 0.62$~GeV for $q=u,d$ (from $I=1$) and
0.45~GeV for $c$. The fitted values of
$\alpha_s^{ft}$ are 1.3 and 0.72 respectively, each determined to
a few per cent accuracy.
The $E_P-E_S$ separation proved to be
quite sensitive to
the strength of the Coulomb potential.
The constant $V_0$ was fixed separately
for each flavor by using the
spin-averaged masses
$E_S^{(I=1)}=0.63$~GeV and $E_S^{(c\bar c)} = 3.07$~GeV
as input.
This required $V_0^{(I=1)}=-1.71$ GeV and $V_0^{(c\bar c)}=-1.17$ GeV.
Since these constant contributions
cancel zero-point energies, they are not physically relevant. One might expect
them
to be roughly flavor independent, however, which can be achieved by increasing
$m_c$ to
1.8 GeV; the effect
on the hybrid spectrum will be discussed subsequently.
For $s\bar s$ we used the $u,d$ parameters and simply increased the quark
mass to $m_s=0.55$ GeV.

The Monte Carlo technique was used to determine masses of
$q\bar q$ and hybrid states
up to $L=3$.
For $L>0$ $q\bar q$ states we used
\begin{equation}
f^{(L,M)}(\theta,\phi) = P_L^{M}(\cos(\theta))\; \cos(M \phi)
\end{equation}
in the guiding wavefunction (38) and the high statistics runs used $M=0$.
For the hybrids we again used the rigid-body angular
wavefunction (41). Tests of node dependence were carried out by varying $M$.
The simulations used
the same statistics as the $\alpha_s=0$ studies of the previous
section, although we found that $\tau_1=5.0$ GeV$^{-1}$ sufficed for
convergence
of
level separations to within the statistical errors.
These errors were typically about $\pm 5$ MeV
for quarkonium
states and $\pm 10$ MeV for hybrids. The guiding wavefunction parameters
used in (34) were $\xi_{q\bar q} = 3 /(2m_q\alpha_s^{ft})$ (to give
an accurate Coulomb wavefunction for S-waves at short distance), and the flux
tube
length scale $\xi$ was optimized numerically for each state. For all $q\bar q$
and
$c\bar c$ states we found that $\xi=1.5$~fm was nearly optimum. For $q\bar q$
hybrids
we found $\xi=1.8$ fm for $\Lambda=1$ and 2.4 fm for $\Lambda=2$. (Note that
the higher
flux tube excitation requires a larger length scale, as expected.) For $c\bar
c$
hybrids we found slightly smaller flux tube length scales,
$\xi=1.6$ fm for $\Lambda=1$ and 2.1 fm for $\Lambda=2$.
The quarkonium levels were again independent of $M$ to within our statistical
erors, but some bias was evident in the hybrids. This bias decreased with
increasing $m_q$ and $m_b$, as expected. The largest bias was found in the
light ${}_1P$ hybrid, for which $E(M=1)-E(M=0)= 57(9)$~MeV, similar to
our findings for $\alpha_s=0$. This fell to
36(7) MeV for charmonium. The corresponding $E(M=2)-E(M=0)$ bias for
${}_1D$ was 24(13) MeV for $u\bar u$ and 18(9) MeV for $c\bar c$. Measurements
with $\pm|M|$ appear to give equivalent results. For this work we average
over measurements with all values of $|M|=0$ to $L$; the
discrepancies given above imply a systematic uncertainty of about
$\pm 30$ MeV for the $u,d$ ${}_1P$ hybrid, $\pm 20$ MeV for the ${}_1P$
$c\bar c$ hybrid, and rather less for the other states. This error could
be reduced in future work through incorporation of improved nodal surfaces.

Our numerical results with the standard parameter set
$(m_q,m_b,\alpha_s^{ft},a) = $ (0.33 GeV, 0.2~GeV, 1.3, 1.0 GeV/fm)
are shown in Fig.5.
The predicted $D$-wave $q\bar q$
mass of 1.66(1)~GeV is quite reasonable, given the
well-established $D$-wave candidates
$\rho_3(1690), \omega_3(1670)$ and $\pi_2(1670)$. The $F$-wave $q\bar q$
multiplet is
predicted to lie at 2.03(2) GeV, in good agreement with the
$a_4(2040), a_3(2050)$ and $f_4(2050)$.
The lightest hybrid multiplet, which has $\Lambda=1$ and $L=1$
($_\Lambda L ={}_1P$ in our
notation), is at 1.90 GeV with these
parameters. This is identical to the Isgur-Merlin-Paton prediction of
1.9 GeV \cite{MP1,IP2}. Since we are using different versions of the flux tube
model this agreement is somewhat fortuitous, although we will show that our
result is rather insensitive to parameter variations.

In view of the interest in the experimental hybrid candidate at 1775 MeV
\cite{Condo}, which may have exotic
$J^{PC}=1^{-+}$ but $2^{-+}$ and $3^{++}$ are also possible,
we also determined the
mass of the radially-excited $L=2$ $q\bar q$ multiplet, which contains
the first $I=1$ $2^{-+}$ $q\bar q$ level expected above the
$\pi_2(1670)$. (A $3^{++}$ $q\bar q$ state would require $L=3$, and since
this multiplet has well established members near 2.05 GeV we do not
consider this a plausible $q\bar q$ assignment.) For the radial simulation
we
multiplied the $q\bar q$ guiding wavefunction $\psi_g^{(L)}$ in (38)
by $|R-R_0|$, and varied the node radius $R_0$ until the energies determined
by Monte Carlo in the $R>R_0$ and $R<R_0$ regions were equal. This required
$R_0=1.5$~fm and gave an energy of $E_D' \approx 2.3$~GeV,
similar to potential
model expectations \cite{pot}
and far above the 1775 MeV state. This state is thus
very unlikely to be a radially-excited
D-wave $q\bar q$.

We find that the first orbitally excited hybrid multiplet
(${}_1D$) is
at 2.30 GeV, 400 MeV above the
lightest
(${}_1P$)
hybrids. The same numerical result was
found earlier by Merlin \cite{MPhD}
using the adiabatic approximation.
This
${}_1D$
multiplet contains the $J^{PC}$ states
$(1,2,3)^{\pm \mp}$ and $2^{\pm \pm}$, which includes the exotics $1^{-+},
2^{+-}$
and $3^{-+}$.
This level is surprisingly high in mass, since
a small orbital excitation gap has been anticipated for hybrids,
due to the relatively flat hybrid adiabatic potential found by Michael {\it et
al.}
\cite{hylgt} in
heavy-quark lattice gauge theory. We shall see that the orbital excitation gap
is somewhat smaller for $c\bar c$ hybrids in our model, so there is no
serious inconsistency with HQLGT results.
If the experimental hybrid candidates near 1.8 GeV \cite{Condo}
and $1.6$-$2.2$ GeV \cite{f1pi} are confirmed,
it may be useful to search for members of this
${}_1D$ hybrid multiplet near 2.2 GeV (about 0.4 GeV above ${}_1P$).
A sequence of hybrids with higher
orbital excitation is expected in the flux tube model,
although these may be increasingly difficult to
observe due
to small matrix elements with light $q\bar q$ states.

We also determined the mass of the lightest
$\Lambda =2$ hybrid multiplet, ${}_2D$.
These states
are found to be quite high in mass, $\approx 2.75$~GeV, so they should be
irrelevant
for light quark spectroscopy in the 2 GeV mass region.
Merlin and Paton anticipate a lighter two-phonon hybrid multiplet,
near 2.2 GeV in the adiabatic approximation.
In their level the phonon angular momenta cancel
($\Lambda=0$ ``paraphononium"), whereas we have considered $\Lambda=2$
``orthophononium".
These $\Lambda=0$ two-phonon states have conventional
$q\bar q$ quantum numbers, which could complicate their identification.

The sensitivity of hybrid mass predictions to parameter variations is an
important
issue which has received little attention in previous flux tube studies.
To investigate this we sequentially
increased one parameter of the set
$(m_q, m_b, \alpha_s^{ft},a)$ by
$20\% $;
recall that our standard parameter set
(0.33 GeV, 0.2 GeV, 1.3, 1.0 GeV/fm)
gave $(P,D,{}_1P,{}_1D)$ masses of ([1.25](input),1.66,1.90,2.30) GeV. ($V_0$
is always
chosen to give $M_S= (3M_\rho + M_\pi)/4 = 0.63$~GeV.) The variations
of these masses with
parameters (with errors of typically $\pm 0.01$ GeV)
were
\begin{equation}
\Delta (M-M_S) (P,D,{}_1P,{}_1D)\ (GeV) =
\cases{
(-0.01,-0.02,-0.01,-0.02)& ($\Delta m_q/m_q=0.2$), \cr
(-0.01,+0.01,-0.05,-0.03)& ($\Delta m_b/m_b=0.2$), \cr
(+0.07,+0.08,+0.06,+0.09)& ($\Delta \alpha_s^{ft}/\alpha_s^{ft}=0.2$), \cr
(+0.05,+0.11,+0.13,+0.16)& ($\Delta a/a=0.2$). \cr
}
\end{equation}
This leads to several conclusions about the importance of
parameter uncertainties in our flux tube spectrum.
First, the level separations are
evidently quite insensitive to variations in
quark mass. Second, they are sensitive to changes
in $\alpha_s^{ft}$ and $a$, but the known $P$-$S$ and $D$-$S$
$q\bar q$ separations preclude
any large changes in these parameters. In any case the hybrid and $D$-wave
levels
behave similarly under changes in $\alpha_s^{ft}$ and $a$, so
the predicted hybrid to $D$-wave
separation
is quite stable. Finally, it is the bead
mass that leads to the largest uncertainty. The energies do not depend
especially strongly
on this parameter, but the hybrid and $q\bar q$ energy shifts have
opposite signs. (This is more evident in (46) below.).
Unfortunately the $q\bar q$ masses are quite insensitive to $m_b$, so
ideally we would use a hybrid mass to determine $m_b$.
To estimate the range of plausible hybrid masses as we vary $m_b$
we consider the range
$m_b=0.2$-$0.4$ GeV; $0.2$ GeV is our standard value and $0.4$ GeV corresponds
to
a large flux tube length scale (see discussion in Sec.II.B).
Over this range of $m_b$ we find the masses (with square brackets as input
data)
\begin{equation}
(S,P,D,{}_1P,{}_1D)\ \hbox{(GeV)} =
\cases{
([0.63],[1.25],1.66,1.90,2.30)& ($m_b=0.2$ GeV), \cr
([0.63], 1.27 ,1.70,1.78,2.22)& ($m_b=0.4$ GeV). \cr
}
\end{equation}
With rounding to 0.1 GeV accuracy this leads to our final estimate of the
lightest hybrid mass,
\begin{equation}
M({}_1P) = 1.8\hbox{-}1.9  \ \hbox{GeV} \ .
\end{equation}
The first orbitally excited hybrid
${}_1D$
and the first $\Lambda=2$ hybrid
${}_2D$ are expected at about 0.4 GeV and 0.8 GeV above the
${}_1P$ hybrid level respectively.

For $s\bar s$ quarkonia and hybrids we simply increased $m_s$ to 0.55~GeV.
The resulting level splittings were very similar to the results for $u,d$
states.
Using a $P$-wave $s\bar s$ mass of 1.50~GeV as input to fix
$V_0$, our $s\bar s$
results are
\begin{equation}
(S,P,D,{}_1P,{}_1D)\ \hbox{(GeV)} =
(0.87,[1.50],1.88,2.17,2.54) \ \ \ (m_b=0.2 \ \hbox{GeV}).
\end{equation}
The only significant changes noted were a decrease in the $D$-wave
level (relative to $E_S$) of $\Delta (E_D-E_S) = -0.02$~GeV and an increase
in the ${}_1P$ level by $0.03$~GeV. Thus we expect the first $s\bar s$ hybrid
near $M_D(s\bar s) + 0.29$ GeV,
about 50~MeV higher above the $D$-wave level
than we found for the corresponding $u,d$
states.
The dependence on $m_b$ was very similar to that found for $u,d$, so our
final result for the first $s\bar s$ hybrid level ${}_1P$ was 2.1-2.2~GeV.

For charmonium and $c\bar c$ hybrids
with our standard parameters $m_c=1.5$ GeV, $m_b=0.2$ GeV,
$\alpha_s^{ft}=0.72$
and $a=1.0$ GeV/fm
we predict the following levels:

\begin{equation}
(S,P,D,{}_1P,{}_1D)\ \hbox{(GeV)} =
([3.07],[3.52],3.77,4.21,4.48) \ \ \ (m_b=0.2 \ \hbox{GeV}).
\end{equation}
These are displayed in Fig.6.
Note that the predicted $D$-wave $c\bar c$
mass of 3.77 GeV is in good agreement with the experimental $\psi(3770)$.
With these parameters we expect the lightest charmonium hybrid at 4.2 GeV.
The first orbital excitation gap of $c\bar c$ hybrids in HQLGT
was found to be 0.22 GeV by Michael et al. \cite{hylgt} whereas we estimate
0.27 GeV; given the approximations this
does not represent a serious discrepancy, although we shall see below that it
is
a rather stable prediction of this version of the flux tube model.

To test the sensitivity of these results to parameters we again
increased each parameter in turn by
$+20\% $, which gives the mass shifts
\begin{equation}
\Delta (M-M_S) (P,D,{}_1P,{}_1D)\ \hbox{(GeV)} = \cases{
(+0.02,+0.03,+0.04,+0.04)& ($\Delta m_c/m_c=0.2$), \cr
(+0.01,+0.02,-0.05,-0.02)& ($\Delta m_b/m_b=0.2$), \cr
(+0.10,+0.13,+0.07,+0.011)& ($\Delta \alpha_s^{ft}/\alpha_s^{ft}=0.2$), \cr
(+0.04,+0.06,+0.14,+0.14)& ($\Delta a/a=0.2$). \cr
}
\end{equation}
Thus for hybrid charmonium we reach similar conclusions regarding
parameter
uncertainties. The results are quite insensitive to $m_c$; increasing $m_c$
from 1.5 GeV to 1.8 GeV only increases the first hybrid mass by 40 MeV.
Since charm quark
masses from 1.25 GeV (HQLGT, \cite{hylgt}) to 1.77 GeV (flux tube,
\cite{MP1,IP2,MP2}) have been used in the
hybrid literature, it is reassuring to find that
the lightest hybrid mass changes by only about 0.1 GeV over this wide range.
As with light quarks we find that
$a$ and $\alpha_s^{ft}$ strongly affect the hybrid mass spectrum,
however these parameters are
tightly constrained by the known quarkonium spectrum.
The largest uncertainty again comes from $m_b$, which is not very well
determined by the $c\bar c$ spectrum nor by more general theoretical
considerations. To test a wide range of possible values we
again vary $m_b$ over the range
$m_b=0.2$-$0.4$ GeV; with $m_b=0.4$ GeV we find

\begin{equation}
(S,P,D,{}_1P,{}_1D)\ \hbox{(GeV)} =
([3.07],3.54,3.82,4.08,4.37)\ \ \  (m_b=0.4 \ \hbox{GeV}).
\end{equation}
Our final result for the lightest hybrid charmonium mass is thus
\begin{equation}
M({}_1P) = 4.1\hbox{-}4.2 \ \hbox{GeV}   \ ,
\end{equation}
and for charmonium we expect the orbital
(${}_1D$)
and doubly-excited
(${}_2D$) hybrids about 0.3~GeV and 0.7-0.8 GeV above the
${}_1P$ level respectively.

\section{Phenomenological Implications}

We have studied the fixed-$N$ version of the flux tube model, principally
the $N=1$ case, as a numerically
tractable version of this type of hadron model.

The ability to reproduce the spectrum of conventional quarkonia with
$N = 1$ is of interest in its own right. It suggests
that we have a unified picture of both quark and
flux-tube excitation spectra, thereby generating some confidence in the
predicted hybrid masses. In this final section we
summarize implications of these results.

Our studies suggest that the adiabatic
approximation, used in previous
analyses of hybrid meson masses in the flux tube model,
underestimates the hybrid mass scale. Our conclusions substantiate
previous analytical estimates of
corrections to the adiabatic approximation\cite{MP1,MP2}, and lead to
hybrid masses that are $\approx 0.1$~GeV above the predictions
of quenched heavy-quark lattice QCD, but are consistent with these lattice
results given their
estimated corrections to the quenched approximation.

In contrast to the light quark sector, in which flavor mixing in non-exotics
may be important and the $q\bar q$ spectrum itself is rather controversial,
in heavy-quark systems the $Q\bar Q$ spectroscopy is relatively
straightforward and special opportunities ensue for the detection of hybrids.
Our results support the expectation that heavy
hybrids, $H_Q$, appear at masses of
\begin{equation}
M(H_Q) \approx M_0(Q\bar{Q}) +1\; {\rm GeV}\ .
\end{equation}
An important feature in heavy $Q\bar{Q}$ spectroscopy is the existence
of narrow states spanning a mass range from $\approx M_0(Q\bar{Q})$
through  $\approx 1$ GeV up to the two-body open-flavor
threshold (i.e. $\psi$ to $D\bar{D}$ or $\Upsilon$ to $B\bar{B}$).
So for charmonium hybrids, for example, one anticipates $H_c$ states
in the resonance region not far above the open charm threshold
of 3.73 GeV.
In our simulations we actually find the first charmonium hybrids at
$M(H_c) = $4.1-4.2 GeV.

Such a prediction is particularly exciting.
Charmonium spectroscopy is rather well understood up to about 3.8 GeV,
so searches for unusual states should be straightforward near this mass.
Since
only a few open charm channels occur below 4.3 GeV,
for a considerable range of hybrid masses one might anticipate rather narrow
hybrid resonances. This possibility receives additional support from
the flux tube model
\cite{IKP,CP}, which predicts that
the dominant two body decay modes of the lowest lying hybrids
are an $L=0$ and $L=1$ $q\bar q$ meson pair.
These $S+P$ thresholds are rather high in mass,
about 4.3 GeV for $c\bar c$ hybrids and 11.0 GeV for
$b\bar b$ hybrids.
The possibility that relatively narrow hybrid charmonium states may exist
within this $3.8$-$4.3$ GeV window provides an
exciting opportunity for $e^+e^-$ facilities such as BEPC, KEK
and a Tau-Charm Factory. If there
are indeed hybrids at these masses, one expects that they should
be produced copiously by gluon fragmentation at large
momentum transfers, for example at the Tevatron.
Detection of the $\psi$ or $\psi(3685)$ as a signature of
hadronic cascade decays of metastable
hybrid charmonia
has been discussed in ref \cite{C94}. (A double cascade from the $c\bar c$
continuum to a hybrid and thence to $c\bar c$ was proposed for a
Tau-Charm Factory by D.V.Bugg, see ref \cite{TCF}.)
In practice the usefulness of cascade
decays in hybrid searches
will depend on their branching fractions to conventional quarkonia.

Determination of the production and decay characteristics
of hybrid states is beyond the scope of this study, but
we note in passing that progress
in this area has been made recently by analytical modelling of
flux tube excitations\cite{CP,CP1}. In these references
the decay amplitudes of some recently discovered
$1^{-+},0^{-+},1^{--}$ and $2^{-+}$ $u,d$-flavored mesons were found
to be in good agreement with the predicted properties of hybrid mesons,
so the flux tube model may be a useful guide to strong decay modes as well
as masses.
Widths of the hybrid charmonia
calculated in this model support the suggestion that
some of these $c\bar c$ hybrids are likely to be narrow.

The production of $1^{--}$ charmonium vector hybrids seems especially
promising.
As the flux tube has an orbital excitation about the $q\bar q$ axis, and the
$q\bar q$ themselves have an effective centrifugal barrier due to the
flux tube angular momentum, which suppresses the radial $q\bar q$ wavefunction
at small $r$, we anticipate that the $e^+e^-$ widths $\Gamma_{ee}(V_c)$
should be significantly smaller than those of the conventional $c\bar c$
states $\psi$ and $\psi(3686)$.

In light quark systems
this wavefunction suppression is not dramatic (see for example the
Particle Data Group summary of $V \rightarrow e^+e^-$ \cite{PDG}
for $L=0$ and $L=2$ $q\bar q$ states following
the analyses of refs\cite{donnachie}), so we anticipate
a significant light hybrid leptonic width $\Gamma_{ee}(\rho_g ) $.  The
principal
difficulty here may lie in distinguishing between
light conventional and hybrid vector states unambiguously.
The recent analyses of the light vector sector by Donnachie and Kalashnikova
\cite{dk} actually do support
the presence of additional vector states, some of
which they suggest may be hybrids.

The recent studies of hybrid decays in the flux tube
model\cite{CP,CP1} may allow tests of these possible
light vector hybrids. Since the $q\bar{q}$ pair in $V_g$ has
$S_{q\bar q}=0$,
whereas conventional $q\bar q$
vector states (either $^3S_1$ or $^3D_1$) have
$S_{q\bar q}=1$, there are characteristic selection rules for decays that
discriminate between these spin-singlet and triplet states.
In particular, if the $q\bar{q}$ are in a spin singlet (as in the $V_g$
vector hybrid case)
then the flux tube decay model
forbids decays into final states of two spin singlet mesons.

For $J^{PC}=1^{--}$ states this selection rule
distinguishes rather clearly between conventional
 and hybrid vector mesons. It implies that in the decays of
a light $\rho_g$ hybrid
$\rho_g \not\rightarrow \pi h_1$, although $\rho_g\to \pi a_1$ is allowed.
Analogously, $\omega_g\not\rightarrow\pi b_1$ for hybrid $1^{--}$
$\omega_g$ decays; this
is opposite to the case of conventional $^3L_1$ $q\bar q$ mesons, for which
the
$\pi a_1$ channel is suppressed relative to $\pi h_1$ or $\pi b_1$
\cite{busetto,kokoski87}. The extensive analysis of data
in ref \cite{donnachie} revealed the clear presence of a
$\rho(1450)$\cite{PDG}
with a strong $\pi a_1$ mode but no evidence for $\pi h_1$,
in accord with expectations for a hybrid.
Furthermore, ref\cite{donnachie}
finds an $\omega (1440)$ with no
evidence for decays into $\pi b_1$,
again in conflict with expectations
for conventional $q\bar{q}$ $(^3S_1$ or $^3D_1$) states but in accord
with predictions for hybrid decays.

The branching fractions reported for the $\rho(1450)$\cite{donnachie}
(see also \cite{CP1})
suggest that there may be mixing between
$\rho_g$ and radial $\rho$ basis states in
this region. If these hybrid states near 1.5 GeV are confirmed,
this mixing may explain the low mass relative
to the $1.8$-$2.0$ GeV typical of other hybrid candidates.
There may also be significant spin-dependent mass
shifts in hybrids that were not incorporated in the present study,
which reduce spin-singlet masses
(such as $V_g$) relative to the spin triplet states
($0^{-+/-+},1^{-+/+-},2^{-+/+-}$).
To test this possibility,  analogous experimental investigations of
$1^{--}$ hybrid
charmonia in $e^+e^-$ would be very useful.
In contrast, in $b\bar{b}$ systems the suppressed
wavefunction at contact is expected to make $H_b$
hybrids essentially absent in $e^+e^-$ annihilation. For this reason
the
charmonium system may
be optimal for hybrid searches;
conventional $c\bar{c}$ spectroscopy
is reasonably well established, and since the $D$-wave coupling
$\Gamma_{ee}(\psi (3770) )$
is not negligible, it may be possible to
observe a moderately suppressed $V_c$ vector hybrid signal in $e^+e^-$
annihilation at a Tau Charm Factory\cite{TCF}.
Diffractive
photoproduction
of charmonium hybrids,
$\gamma^* P \to X P$, may also be possible,
for example at HERA.

If the mass of the $V_c$ is indeed below or near
4.3 GeV ($D^{**}\bar D$ threshold), then hadronic cascades to
conventional charmonium states,
in particular the $\psi(3097)$ and $\psi(3685)$, may be important and
could
provide a good tag\cite{C94}. The E835 experiment at Fermilab may
be able to observe production of hybrid charmonium through hadronic
cascade decays to $\psi \pi \pi$ and $\psi \eta$.

For hybrids which lie above $D^{**}\bar D$ threshold
heavy quark symmetry or detailed decay models may be used to distinguish
the spin singlet $H_c$ from the spin triplet $\psi$ states through
their decay systematics. More
detailed theoretical study on this and related questions is now warranted.

To summarize, we find that heavy-quark hybrids in the flux tube
model lie below $S+P$
thresholds,
and for hybrid charmonium this implies
that the lightest states should
have rather narrow widths. We anticipate that production by gluon jets
may be particularly promising and for this case some quantitative
estimates already exist\cite{C94} based on the masses found here.

In conclusion, we find the lightest hybrid masses in the flux tube model to be
$M(H_{u,d}) = 1.8$-$1.9$ GeV and
$M(H_c) = 4.1$-$4.2$ GeV. These results, combined with
recent detailed studies of hybrid decay modes
\cite{CP,CP1},
provide a clear set of theoretical predictions for hybrids
for comparison with experiment.

\acknowledgements

We would like to acknowledge useful discussions or communications
with E.S.Ackleh, G.Condo, J.Govaerts, N.Isgur, S.Narison, P.R.Page and J.Paton.
This research was sponsored in part by
the European Community  Human Mobility program EURODAFNE,
contract number CHRX-CT92-0026; and the United States Department
of Energy under contract DE-AC05-840R21400 with
Martin Marietta Energy Systems Inc. at Oak Ridge National Laboratory.

\begin{figure}
\caption{
Ground state and first hybrid adiabatic potentials and their difference,
for $N=1$. Solid lines are exact and dashed lines are the small oscillation
approximation. String tension $a$=1.0 GeV/fm, bead mass $m_b=0.2$ GeV.
}
\end{figure}

\begin{figure}
\caption{
Hybrid potential gap $E_1(R)-E_0(R)$
for $N=1$ and $N=2$. Plotting conventions and parameters as in Fig.1; points
are
Monte Carlo.
}
\end{figure}

\begin{figure}
\caption{
An $N=1$ quark, antiquark and flux-tube bead, showing the $q\bar q$-axis
angles $\theta$ and $\phi$ and
the rigid-body rotation angle $\phi_b$ relative to the reference
configuration.}
\end{figure}

\begin{figure}
\caption{
Energies of the lightest $L=1,2$ $q\bar q$ and
${}_\Lambda L
= {}_1P$ hybrid states
relative to $E_0=E_S$ for $N=1$.
Lines show the adiabatic approximation and
the points are Monte Carlo, $M=0$ (open) and $M=L$ (plus).
Parameters $m_b=0.2$ GeV, $a=1.0$ GeV/fm, $\alpha_s=0$.
}
\end{figure}

\begin{figure}
\caption{
The lightest $L=0$-$3$ $q\bar q$ ($q=u,d$) and
${}_\Lambda L
= {}_1P$, ${}_1D$ and ${}_2D$ hybrid masses from Monte Carlo
with physical parameters,
$m_q=0.33$ GeV, $m_b=0.2$ GeV, $a=1.0$ GeV/fm, $\alpha_s^{ft}=1.3$.
Square brackets denote masses used as input.
}
\end{figure}

\begin{figure}
\caption{
Charmonium $c\bar c$ and hybrid masses, legend as in Fig.5.
Parameters modified for charmonium are $m_c=1.5$ GeV and $\alpha_s^{ft}=0.72$.
}
\end{figure}

\end{document}